# Predicting the propensity for thermally activated β events in metallic glasses via interpretable machine learning

Qi Wang[1*], Jun Ding[2*], Evan Ma[1]

[1] Department of Materials Science and Engineering, Johns Hopkins University, Baltimore, Maryland 21218, United States

[2] Center for Advancing Materials Performance from the Nanoscale (CAMP-Nano), State Key Laboratory for Mechanical Behavior of Materials, Xi'an Jiaotong University, Xi'an 710049, China

[*] Correspondence to qiwang.mse@gmail.com (Q. W.), dingsn@xjtu.edu.cn (J. D.)

# ABSTRACT

The elementary excitations in metallic glasses (MGs), i.e., β processes that involve hopping between nearby sub-basins, underlie many unusual properties of the amorphous alloys. A high-efficacy prediction of the propensity for those activated processes from solely the atomic positions, however, has remained a daunting challenge. Recently, employing well-designed site environment descriptors and machine learning (ML), notable progress has been made in predicting the propensity for stress-activated β processes (i.e., shear transformations) from the static structure. However, the complex tensorial stress field and direction-dependent activation could induce non-trivial noises in the data, limiting the accuracy of the structure-property mapping learned. Here, we focus on the thermally activated elementary excitations and generate high-quality data in several Cu-Zr MGs, allowing quantitative mapping of the potential energy landscape. After fingerprinting the atomic environment with short- and medium-range interstice distribution, ML can identify the atoms with strong resistance or high compliance to thermal activation, at an unprecedented accuracy over ML models for stress-driven activation events. Interestingly, a quantitative "between-task" transferring test reveals that our learnt model can also generalize to predict the propensity of shear transformation. Our dataset is potentially useful for benchmarking future ML models on structure-property relationships in MGs.

# INTRODUCTION

Metallic glasses (MGs), as a unique class of amorphous materials, exhibit a high atomic packing density with pronounced topological and chemical short-to-medium range order[1–4]. The complex local structures have been demonstrated to have a profound influence on the properties of MGs[5]. In essence, many properties of MGs can be depicted in terms of excursions in the potential energy landscape (PEL)[6–8], which is a multidimensional configurational space with corresponding local energy minima separated by barriers. In the PEL picture, elementary excitations upon external stimuli (e.g., thermal or mechanical) are associated with the β processes, which correspond to the hopping between nearby local minima, i.e., sub-basins inside a deep PEL megabasin[9]. Elementary excitations have been correlated with many properties[10-12], including local plastic deformation[13,14], diffusion mediated by atomic hopping[15], as well as structural relaxation (local energy minimization in the direction towards the bottom of the basin) or rejuvenation (to a higher-energy local minimum)[16].

It remains as a long-standing challenge to unravel the role of static structure in controlling the elementary excitations in MGs: is there a structural indicator that can be tapped into to predict how resistant or compliant different local regions are to externally stimulated activation? Over the past several decades, many efforts have been devoted to addressing this critical question. Recently, the emerging machine learning (ML) technique, based on well-crafted representations of atomic environment, has been proven to be promising for establishing atomic-level structure-property relationships in liquids and glasses[17–24]. For example, Schoenholz et al.[17] studied L-J model liquids and utilized ML to derive a structural parameter called "softness", which was found to correlate well with the particle's propensity for hopping, reflecting its susceptibility to β relaxation of liquids[10]. Below the glass transition temperature, however, metallic liquids become frozen into glass solids and the timescale of the glass dynamics becomes very long, well beyond the capability of atomistic (e.g. molecular dynamics) simulations. We therefore have to resort to local perturbation methods, to activate the local group of atoms into excited states by stress or thermal stimulus, as a probe into the susceptibility to elementary excitations. Several recent ML studies have focused on quantitatively gauging how local environment influences the propensity for stress-activated β processes (i.e., shear transformations) in MGs[18-20]. For example, pioneering works of Cubuk et al.[18] performed ML on disordered materials such as L-J glasses and granular systems and showed that radial and bond-angle distribution information can be used to identify atoms with high propensity to shear transformation. Wang et al.[19] developed interstice distribution as a new local structural representation for MGs, which is proven to be robust in predicting plastic sites of several MGs and has advantages in generalizing between compositions even chemical systems. However, the accuracy achieved in these attempts is not yet sufficiently high, and the reported scoring metric, e.g. recall or area under receiver operating characteristic curve (AUC-ROC), is typically below 80%. One reason for this is that the elementary excitations upon shear transformations are complicated by the non-uniform tensorial stress field in the solid under deformation, as well as the dependence of activation on loading conditions (e.g., loading mode and direction)[25,26]. If not properly dealt with, these would introduce non-trivial noises in the accrued data and influence adversely the quality of the learnt structure-property relations.

This problem, however, subsides when dealing with the thermally induced elementary excitations in MGs. For instance, here we use activation-relaxation technique (ART)[27,28] to probe the propensity for thermal activation of each atom in MGs (see schematic description in Fig. 1a, will be discussed later). These activated processes are not subject to internal non-uniform stresses, and can be well converged by averaging over a considerable number of activation pathways, significantly reducing data noises. Meanwhile, the Gaussian-like distribution of thermal activation energetics (to be shown later) can well identify atoms at both the hard and soft ends, corresponding to locally favored and unfavored motifs, respectively. This avoids the problem associated with common stress activation indicator (e.g., non-affine displacement or von Mises strain), which often exhibits a skewed "long-tail" distribution[29] and the resolution at the hard end is much lower than that on the soft side. Moreover, thermally activated events are comparable in their energetics, at least for some MGs that are based on some common (or similar) elements, even when they are of different composition or processing history; as such these multiple "datasets" can be combined to facilitate the ML identification of the structural underpinning in more general terms.

In this work, we develop ML models to predict the propensity of thermally activated elementary excitation, from the atomic environment of the static MG structure. We systematically probe the activation energies in six MGs, including $Cu_{64}Zr_{36}$ prepared using different quenching rates, as well as $Cu_{50}Zr_{50}$ and $Cu_{80}Zr_{20}$, using ART[27,28]. The activation energy around each atom is calculated, and ensemble-averaged over 50 activation trials, to indicate its susceptibility to excitation. We then combine the data from the six MGs into a wider activation energy spectrum (Fig. 1b) and use ML to identify those atoms with strong resistance or high compliance to activation. By fingerprinting the atomic site environment with a recently proposed interstice distribution representation[19], we find that ML can reliably identify atoms with the highest 5% and lowest 5% activation energy, reaching an area under receiver operating characteristic curve (AUC-ROC) of 0.942 and 0.888, respectively. Such accuracies are considerably better than that in previous ML prediction of the propensity for stress-driven shear transformations[18,19]. We also identify descriptors that are critical to ML decision, and interestingly, most of them turn out to be medium-range order features. Finally, we conduct quantitative "between-task" transferring tests and show that our learnt model can be used to predict the propensity for shear transformation as well. This ML work highlights the predictive power of local static structure to quantitatively connect with β processes in MGs.

## RESULTS

### Energy barriers for thermally activated β processes

We employ molecular dynamics (MD) simulation to prepare six Cu-Zr model MG samples: i) different compositions yet with the same cooling rate ($Cu_{50}Zr_{50}$, $Cu_{64}Zr_{36}$, $Cu_{80}Zr_{20}$ quenched from liquid at $10^{10}$ K/s), and ii) same composition but with different cooling rates ($Cu_{64}Zr_{36}$ MGs with the quenching rates of $10^9$ to $10^{12}$ K/s) (see Methods for simulation details). We then apply ART to probe the energy barrier for thermally-activated events[27,28]. Around each atom in those MGs, we initiate 50 independent activation events along random activation pathways (illustrated by the dashed red lines in Fig. 1a, see Methods for more details). The ensemble-averaged activation energy, $E_{act}$, can then be defined as the average energy difference between the saddle point and the initial state,

$$E_{act} = \left\langle E_{saddle} - E_{initial} \right\rangle \quad (1)$$

The average value of 50 independent activations around each atom are sufficient to achieve a converged $E_{act}$, which contains key statistical information for thermal excitations on each local region (including center atom and its neighbors).

Figure 1b shows the distribution of $E_{act}$ in the six MGs. The dashed vertical line denotes the percentile 50% (median) of $E_{act}$ in each MG. The wide spread of $E_{act}$ signifies a large degree of structural and property heterogeneity in each glass. As mentioned in Introduction, the Gaussian-like distribution of $E_{act}$ observed is very different from that for stress-activated event, where a "long-tail" distribution is often observed in the stress activation indicator (e.g., non-affine displacement or von Mises strain)[29]. The $E_{act}$ spectrum clearly depends on the MG composition or quenching rate. Next, we merge the $E_{act}$ data of the six MGs into a more comprehensive $E_{act}$ spectrum (Fig. 1b). The combined spectrum markedly increases the variety of local environments surveyed, far beyond what is present in a single MG. Later, we will feed this combined dataset to ML and test if ML is capable of mapping out the characteristic atoms at both the high $E_{act}$ (hard) and low $E_{act}$ (soft) ends of these Cu-Zr MGs.

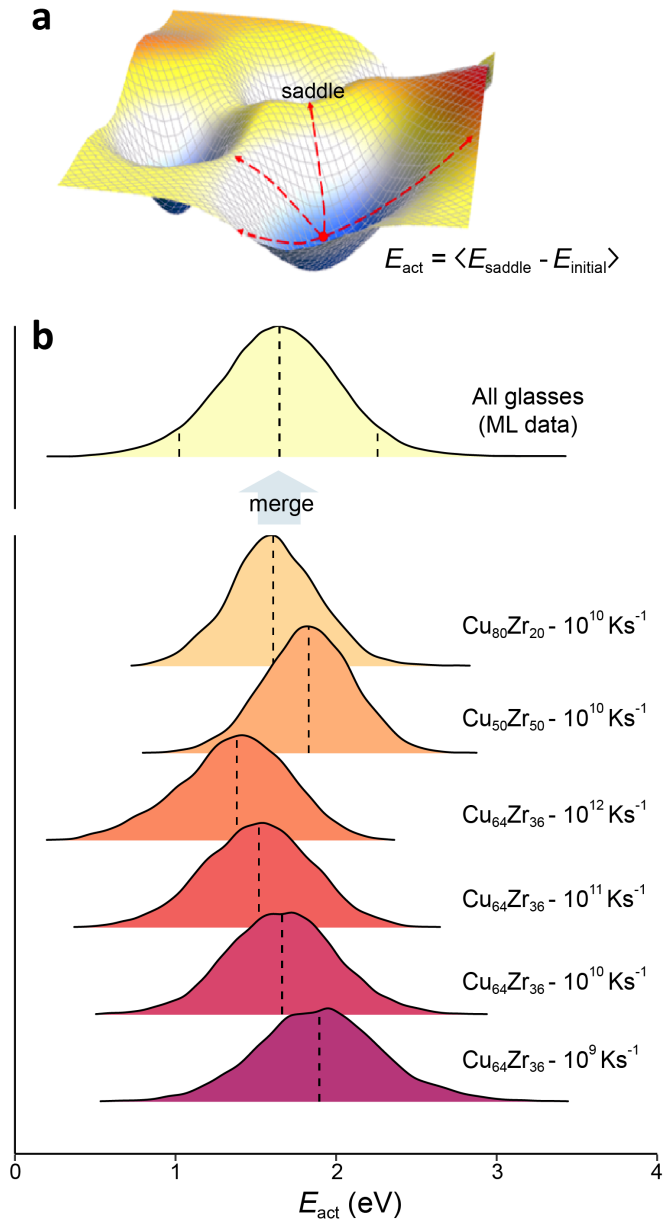

**Figure 1 | Thermally activated events in Cu-Zr metallic glasses.** (a) Schematic description of the β-process in the context of potential energy landscape (PEL). Red dashes illustrate several activated pathways from a local minimum. In practice, we initiate 50 independent events around each atom along random activation pathways using ART and extract an ensemble-averaged activation energy $E_{act}$ for the atom. (b) Distribution of $E_{act}$ in the six model glasses as well as their combined $E_{act}$ spectrum. The median (quantile 50%) and quantiles 5% and 95% are marked as vertical dashed lines in the combined $E_{act}$ spectrum, and the median (quantile 50%) is marked in the spectrum of each model glass.

## Connecting activation barriers with local atomic environment

We make use of a set of interstice distribution descriptors to represent the local atomic environment[19]. The basic fingerprinting procedure is to extract groups of bonds, facets and tetrahedra from the coordination polyhedron of an atom, and then featurize the distribution of interstitial spaces present in these bond, facet and tetrahedron groups. A simple treatment of representing a distribution is to derive typical statistics (such as minimum, mean, maximum and standard-deviation) of the interstitial spaces present. The characterization of bond, facet and tetrahedron interstices can include 2-body, 3-body and 4-fold correlations, respectively, in the nearest-neighbor short-range-order (SRO) signatures. The SRO signatures will be further "coarse-grained" to derive statistics among their neighbors. Such "coarse-grained" signatures are a representation of medium-range-order (MRO), with a length scale of ~4 – 6 Å, which is the next-level structural organization beyond the SRO. Upon implementation, the interstice representation contains 80 descriptors, 16 SRO and 64 MRO. The codes for this representation can be publicly accessed in *amlearn*[19] (https://github.com/Qi-max/amlearn) and *matminer*[30] (https://github.com/hackingmaterials/matminer). This representation has been demonstrated to be highly predictive, interpretable and generalizable in a range of MGs[19].

After featurizing all atoms in the six MGs, we feed the data to a scalable tree boosting ML algorithm, XGBOOST[31]. XGBOOST implements a parallel tree boosting algorithm that is proven to be very efficient and robust in various cases. We train two sets of XGBOOST classifiers to identify the highest 5% and the lowest 5% $E_{act}$ atoms, respectively, in the combined dataset merged from six MGs (Fig. 1b). We have tested that varying the threshold from 3% to 10% gives similar results, and in general, the smaller the fraction, the better the ML score (i.e., the easier for ML to identify). As we are dealing with an imbalanced dataset, we do random equal undersampling three times to create three data samples, each with 3,000 positive class atoms (the highest or lowest $E_{act}$ atoms) and 3,000 negative class atoms. We then perform 5-fold cross-validation on each of the data samples, and average the predictions on the test sets (i.e. averaged over 5 × 3 = 15 test sets). The repeated undersampling procedure is very useful for reducing the variance introduced by data undersampling.

We use the area under the receiver operating characteristic curve (AUC-ROC) as the scoring metric of the classifiers. The ROC curve characterizes the tradeoff between the true positive rate (TPR) and negative positive rate (FPR)[32]. TPR is also known as recall or sensitivity: TPR = TP / (TP + FN), where TP and FN are short for true positive and false negative, respectively. FPR is the false-alarm rate: FPR = FP / (FP + TN), where FP and TN stand for false positive and true negative, respectively. AUC-ROC, measuring the area underneath the ROC curve, is a widely used metric to evaluate a classifier[32]. By definition, an AUC-ROC of 0.5 indicates a random chance level, 1.0 signifies perfect classification accuracy, and the higher the AUC-ROC, the better the model is at distinguishing the classes. Figure 2a presents the ROC curve and its AUC in classifying the highest and lowest 5% $E_{act}$ atoms, respectively. For simplicity, these two ML problems are referred to as "H-$E_{act}$" and "L-$E_{act}$" hereafter. We see that the XGBOOST model trained from interstice distribution can well distinguish the high $E_{act}$ atoms from the rest of the glass, reaching a very high AUC-ROC of 0.942. These high $E_{act}$ atoms are particularly resistant to thermal activation and "pin" the local rearrangement. While there is an increased ambiguity in classifying the lowest $E_{act}$ atoms, the AUC-ROC is also high (0.888), indicating there is also significant structural contrast at the soft end. One can directly observe from the ROC curve the TPR and FPR values at various probability thresholds for designating the classes.

Besides outputting a "label" (0 or 1) to predict whether an atom belongs to a class or not, XGBOOST (and many other ML algorithms) can also give continuous probability estimates, in the range of [0, 1], to reveal the confidence level of predictions. The probabilities can reveal the uncertainty of prediction, allow some flexibility in using the model, and provide a more nuanced way to assess the model. However, raw class probabilities from nonlinear ML algorithms are often not well calibrated and should be carefully checked before interpretation. Specifically, if the predicted probabilities match the "real" class probabilities, such probabilities are referred to as calibrated. For instance, when the positive class probability of some data points is 0.70, ideally these points should indeed have a probability of 0.70 to be positive. This ideal calibration occasion is illustrated by the diagonal line in Fig. 2b. In this work, we employ a post-training calibration method called isotonic regression[33] to improve the calibration performance of our probability estimates. As seen in Fig. 2b, the calibration curves of $p_h$ and $p_l$, i.e. the probability estimates from models obtained in the "H-$E_{act}$" and "L-$E_{act}$" ML problem, respectively, are both demonstrated to be close to perfect calibration. The area between the calibration curve and perfect calibration line, as a measure of miscalibration, is very low in both cases (Fig. 2b). Thus, our machine-learnt probability estimates can well reflect the real class probabilities and warrant further interpretation.

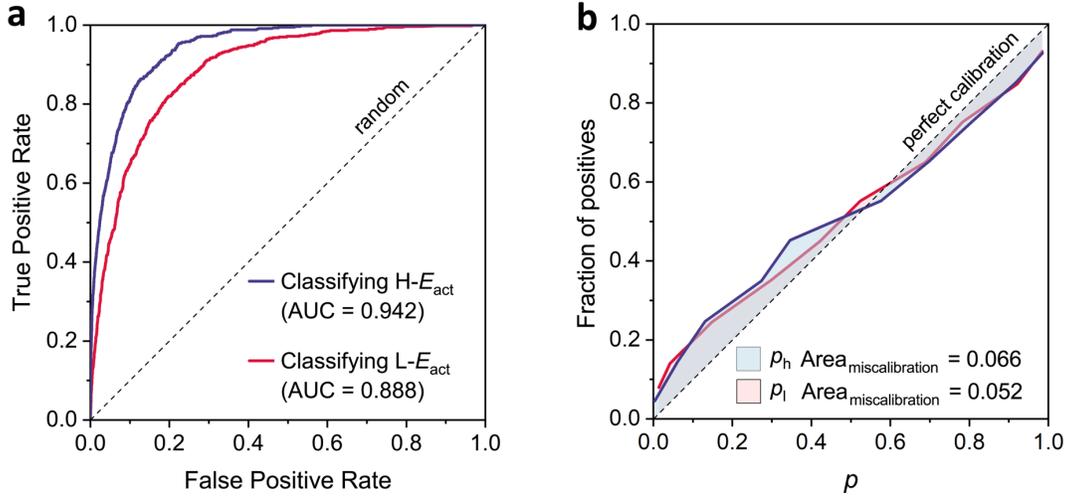

**Figure 2 | Predicting the heterogeneity of thermally activated events**. (a) Receiver operating characteristic (ROC) curve and area under curve (AUC) in classifying atoms showing the highest 5% (H-$E_{act}$ problem) or lowest 5% activation energy (L-$E_{act}$ problem). The dashed line marks random case. (b) Near-perfect calibration of the ML-evaluated class probability estimates, that is, $p_h$ from H-$E_{act}$ and $p_l$ from L-$E_{act}$.

We proceed to look more into the distributions of the ML-evaluated class probability estimates, that is, $p_h$ from H-$E_{act}$ and $p_l$ from L-$E_{act}$. Figure 3a and 3b present the overall $p_h$ and $p_l$ distributions in the six MGs as well as the variation of $E_{act}$ with $p_h$ and $p_l$. A wide distribution of $p_h$ and $p_l$ is observed, suggesting a large degree of heterogeneity inside the MGs. $p_h$ has a larger proportion of atoms close to 0 and 1, again indicating that ML is more confident at distinguishing the high $E_{act}$ atoms. A strong dependence of $E_{act}$ on $p_h$ and $p_l$ is observed, that is, positively correlated with $p_h$ and negatively correlated with $p_l$, demonstrating the feasibility of $p_h$ and $p_l$ serving as indicators of the thermal activation propensity (Fig. 3b). We further visualize the distribution of $E_{act}$, $p_h$ and $p_l$ in a model $Cu_{64}Zr_{36}$ glass to allow atomic-scale scrutinization (Fig. 3c). For simplicity, only atoms with $p_h$ or $p_l$ > 0.50 are highlighted in the $p_h$ and $p_l$ maps: ML predicts that the probability of these atoms having highest 5% $E_{act}$ or lowest 5% $E_{act}$ is greater than 0.50; If setting a class threshold as 0.50, these $p_h$ or $p_l$ > 0.50 atoms would be classified as the high or low $E_{act}$ class, respectively. A good correspondence can be seen between the high $E_{act}$ atoms and high $p_h$ atoms, as well as between the low $E_{act}$ atoms and high $p_l$ atoms. As reflected by the relatively lower prediction score in the L-$E_{act}$ task, there are more false positive atoms (high $p_l$ yet high $E_{act}$) and false negative atoms (low $p_l$ yet low $E_{act}$) in predicting the low $E_{act}$ atoms, but still the prediction quality is sufficiently good. These results reveal that a solid relationship between local structure and thermal activation propensity can be established by combining interstice features and ML.

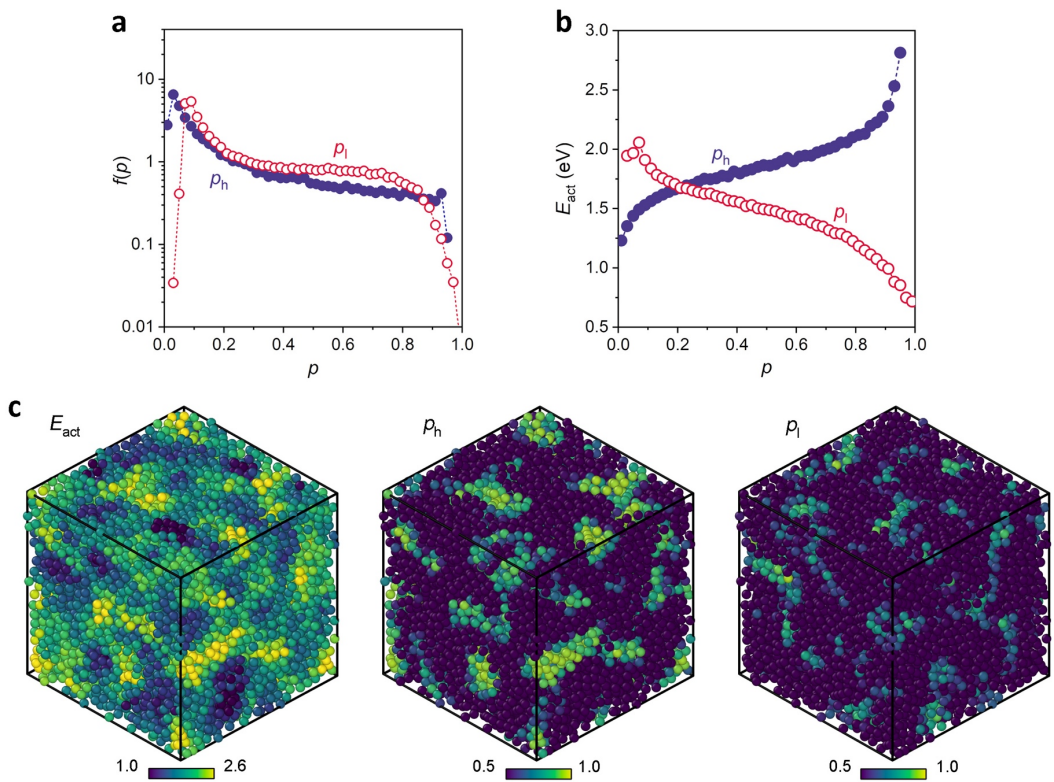

**Figure 3 | ML-evaluated class probability.** (a) Probability density distribution $f(p)$ of $p_h$ and $p_l$ in the combined MG dataset. (b) Strong dependence of activation energy $E_{act}$ on $p_h$ and $p_l$. (c) Distribution of $E_{act}$, $p_h$ and $p_l$ in a $Cu_{64}Zr_{36}$ - $10^9$ K/s MG. For simplicity, only the atoms with $p_h$ or $p_l$ > 0.50 are highlighted in (c). The high and low $E_{act}$ atoms correspond well to the high $p_h$ and high $p_l$ atoms predicted by ML.

## Comparison with ML models employing other feature representations

Next, we compare our ML results based on interstice features with those fitted from several other representations. Here we consider a total of five pure structural representations and three physical signatures for comparison (Table 1). To guarantee a fair comparison, training is performed on the same data samples and same cross-validation splits. We train XGBOOST[31] and SVM[34] models with various hyperparameters and extract the best scores for each representation. Most of the presented scores are from XGBOOST, while the best scores of the radial symmetry functions are from SVM (Table 1). The detailed ROC curves can be found in Supplementary Figs. 1-8. Besides, an additional feature indicating whether the atom is Cu (0) or Zr (1) is added to each representation to help ML decisions. This is very helpful for representations that cannot well distinguish the atom types from the features themselves.

**Table 1 |** ML results in classifying the highest 5% (denoted as H-$E_{act}$ problem) and lowest 5% $E_{act}$ atoms (L-$E_{act}$ problem) of the combined $E_{act}$ spectrum merged from six MGs (Fig. 1b) using different feature representations. The area under receiver operating characteristic curve (AUC-ROC) on the test set is used as the scoring metric. The reported AUC-ROC is averaged from three times of undersampling and fivefold cross-validation on each sampled data.

| Representation | Feature number* | H-$E_{act}$ | L-$E_{act}$ |
| --- | --- | --- | --- |
| Interstice distribution | 80 | 0.942 | 0.888 |
| <0,0,12,0,0> + <0,0,12,4,0> (baseline 1) | 2 | 0.673 | 0.557 |
| Voronoi index (baseline 2) | 5 | 0.750 | 0.628 |
| A group of SRO features | 18 | 0.807 | 0.634 |
| SRO + MRO features | 72 | 0.908 | 0.801 |
| Radial symmetry functions | 70 | 0.909 | 0.774 |
| Flexibility volume, $V_{flex}$ | 1 | 0.845 | 0.784 |
| Atomic shear moduli, $G$ | 1 | 0.690 | 0.629 |
| Coarse-grained $G$ | 1 | 0.736 | 0.674 |

* Except for Interstice distribution, an additional feature indicating whether the atom is Cu (0) or Zr (1) is added to each representation to help ML decisions, so technically the feature number should be added with 1. This is very helpful for representations that cannot well distinguish the atom types from the features themselves.

We start with two "baseline" models built with: i) two one-hot-encoded (0 or 1) variables designating whether the nearest-neighbors around an atom form a <0, 0, 12, 0, 0> (<0, 0, 12, 0> if omitting occasional facets with >6 edges) or <0, 0, 12, 4, 0> Voronoi polyhedron or not; ii) five integer Voronoi indices ($n_3$, $n_4$, $n_5$, $n_6$, and $n_{>6}$), where $n_x$ represents the number of $x$-edged facets in the Voronoi polyhedron[35]. Many studies revealed that the Cu-centered <0, 0, 12, 0, 0> icosahedra and Zr-centered <0, 0, 12, 4, 0> polyhedra are among the most stable motifs in Cu-rich Cu-Zr MGs[36,37]. In this work, ~21.4% Cu atoms of the six MGs are surrounded by icosahedra and 6.0% Zr atoms are <0, 0, 12, 4, 0>. A baseline model can then be simply predicting the icosahedra and <0, 0, 12, 4, 0> atoms as high $E_{act}$ atoms and those not as low $E_{act}$. We find that the AUC-ROC achieved by such baseline model is not very satisfactory, i.e., 0.673 and 0.557 in the H-$E_{act}$ and L-$E_{act}$ task, respectively (Table 1). As seen in the Supplementary Fig. 1a, the TPR (recall) of this baseline model in classifying the highest 5% $E_{act}$ atoms is ~0.48, indicating that indeed only ~0.48 of the highest $E_{act}$ atoms are among the icosahedra and <0, 0, 12, 4, 0> atoms. Not surprisingly, this heuristic model works worse in classifying the lowest $E_{act}$ atoms, as icosahedra and <0, 0, 12, 4, 0> are aimed at prototyping the most stable motifs and not forming those motifs does not necessarily mean that this atom is soft. This results in a large FPR and ultimately a small AUC-ROC of 0.557 in the L-$E_{act}$ task (Supplementary Fig. 1b). As to the second baseline model trained from the Voronoi indices, the prediction is better, with AUC-ROC of 0.750 and 0.628, respectively (Table 1, the ROC curves are presented in Supplementary Fig. 2). We see that by allowing the model to decide from the detailed Voronoi indices instead of several predefined motifs only, the model can capture more subtle structural information and make better decisions in both tasks. These two sets of models are basically based on the well-established Voronoi indices and are relatively simple to set up, forming the baseline models in our tasks, and ideally, any proposed ML models should well outperform the baseline models.

Next, we combine a group of SRO features as the third structural representation, including characteristic motif signatures and Voronoi indices[35] as used in the baseline models, coordination number (CN) within a cutoff distance (4.0 Å) or in a Voronoi polyhedron, Voronoi volume, and bond-orientational order parameters ($q_l$ and $w_l$, where $l$ = 4, 6, 8 and 10)[38]. This representation achieves an AUC-ROC of 0.807 in the H-$E_{act}$ task and 0.634 in the L-$E_{act}$ task (Table 1, see Supplementary Fig. 3 for ROC curves). The inclusion of bond-orientational order features accounts for the increase of AUC-ROC compared with baseline model 2. The L-$E_{act}$ task remains to be a harder task than the H-$E_{act}$ for the structural representation to predict. Beyond SRO, interestingly, if we further augment the SRO features with the coarse-grained MRO features (taking statistics between nearest neighbors[19], as applied in the interstice representation), the predictive ability is greatly enhanced (Table 1, see Supplementary Fig. 4 for ROC curves). This suggests that it is quite important to include MRO for predicting the thermal activation propensity (the importance of MRO will be discussed in more detail later).

As the last pure structural representation, we derive a set of radial symmetry functions[17,18,20,22-24,39,40] to represent the local environment. Radial symmetry functions are part of a structural representation first proposed to fit ML interatomic potential[39,40], and later successfully employed to represent the atomic environment in disordered materials[17,18,20,22-24]. For an atom $i$, the radial symmetry functions are described as,

$$G_\alpha(i;r) = \sum_{j \in \alpha} e^{-(r_{ij}-r)^2/2\sigma^2} \qquad (2)$$

where $\alpha$ represents an atom specie in the system (Cu or Zr). $r_{ij}$ is the distance between atoms $i$ and $j$. $r$ is a variable constant and $\sigma$ is set as 0.2 Å. The sums are taken over all atom $j$ whose distance to $i$ is within a cutoff $R^c$ (6.5 Å). This set of features can be considered as the Gaussian-smoothed partial pair correlation functions at different $r$ values. Here, we vary $r$ from 1.0 to 8.0 Å with a bin size of 0.2 Å (35 bins), generating 35 features for $i - Cu$ and $i - Zr$, respectively. We then use the 70 features as input to train ML models on the same data and cross-validation splits to classify the high $E_{act}$ and low $E_{act}$ atoms. Results are summarized in Table 1 and ROC curves are shown in Supplementary Fig. 5. We see that this representation can well predict the high $E_{act}$ atoms (0.909), yet the score in predicting low $E_{act}$ atoms is relatively lower (0.774). In previous studies, Schoenholz et al.[17] used this radial symmetry function representation to classify atoms with high propensity for hopping (soft end) in L-J liquids and achieved a very high recall of ~90%. The relatively lower accuracy in the current L-$E_{act}$ task (also corresponds to soft end) suggests that identifying atoms susceptible to β relaxation in the solid-state MGs could be harder than that for the parent supercooled liquids, as manifested by that the same set of features achieve a lower score in the former problem. Other possible factors are i) the natural prediction accuracy difference between Cu-Zr MGs described by EAM potential and supercooled liquids described by pairwise L-J potential and ii) the combination of different composition, different quenching rate in a single dataset may increase the ambiguity for the radial symmetry functions.

Finally, we compare the results of pure structural representations with the results of three physical signatures, namely flexibility volume $V_{flex}$[41], atomic and coarse-grained shear moduli $G$[42] (see Methods for details). Table 1 summarizes their prediction scores and the ROC curves are presented in Supplementary Figs. 6-8. These signatures require detailed knowledge of interatomic potentials to calculate and thus are not pure structural representations. Among the physical descriptors, $V_{flex}$ fares much better than atomic or coarse-grained $G$ in correlating with $E_{act}$. We find that the pure structural representations (interstice, SRO + coarse-grained MRO, and radial symmetry functions) are still very competitive compared with these physical signatures (Table 1), further advocating the use of proper structural representation, with the aid of ML, to establish the structure-property relationship in MGs. The interstice distribution features achieve the highest prediction score in both the H-$E_{act}$ and L-$E_{act}$ tasks. Such quantitative benchmarks are important for obtaining a clear picture of the structure-property relations proposed in MGs. We also note that, strictly speaking, the relative performance of each representation is specific to a task. Thus, for a future task of interest, we recommend to conduct some rigorous benchmarking tests to locate the best representation for maximal ML performance.

**Impact of medium-range environment on activated events**

Thus far, we demonstrate that our ML model, employing the interstice features that start from static atomic positions only, can well predict the heterogeneity of thermal-activated elementary excitations in Cu-Zr MGs. We next look into how the ML models make decisions based on the input features.

ML algorithms such as XGBOOST allow quantification of feature importance, which evaluates how each descriptor improves the performance measure, e.g., Gini index for XGBOOST, and thus can be particularly useful for model interpretation. For ease of interpretation, we first remove some highly-linearly-correlated features (Pearson correlation coefficient > 0.70) and then reduce the feature number to 10 by a brute-force recursive feature-elimination procedure: i) train a model with $N$ features and derive the ML performance; ii) iteratively eliminate each of the $N$ features, retrain a ML model with the remaining $N - 1$ features and calculate the performance loss compared to the original model (if any); iii) eliminate the feature with the least performance loss. This is based on that basically, dropping unimportant features should not degrade performance significantly. We recursively repeat the above procedure until the feature dimension is reduced to 10.

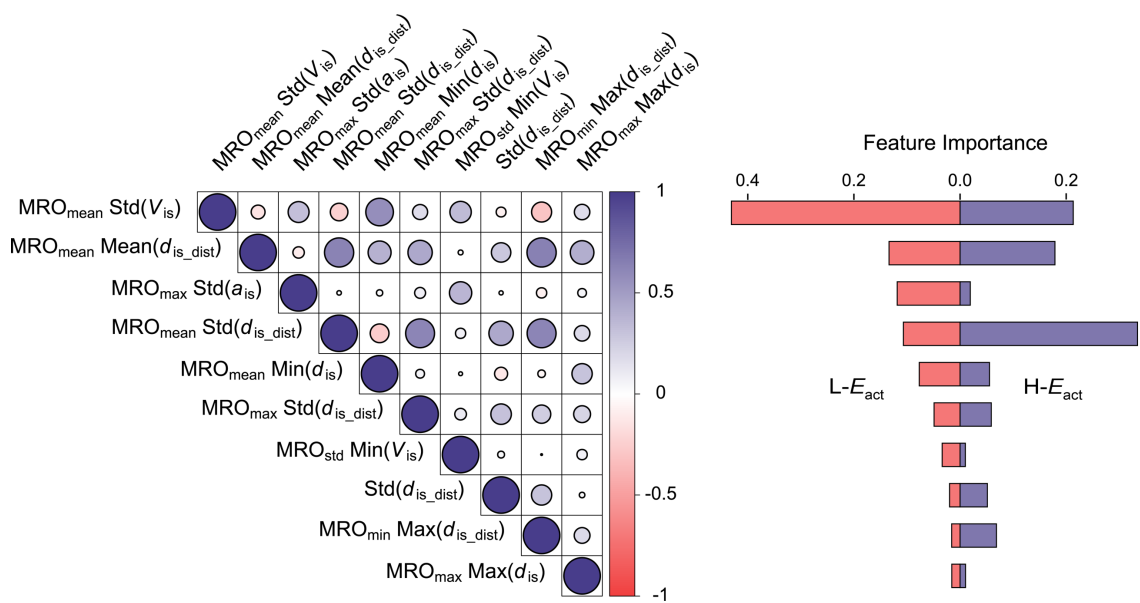

**Figure 4 | Interpreting the ML models.** (a) Pearson correlation coefficient of the ten features that survived the feature reduction. The coefficient value is encoded in the color while the circle radius encodes the absolute coefficient value. $V_{is}$, $a_{is}$ and $d_{is}$ represent the volume, area and distance interstices, respectively, and the symbol before brackets, i.e. Std, Mean, Std, Min, denotes the statistics of these interstices in the nearest-neighbor (SRO) environment; if there is MRO$_{stat}$ in the feature name, this means that the SRO feature has been coarse-grained among neighbors, i.e. taking the statistics, as denoted by the subscript of MRO, among neighbors. (b) Feature importance of the ML models trained in the H-$E_{act}$ and L-$E_{act}$ tasks. The feature importance is averaged over models obtained from the three times of undersampling and five-fold cross-validation in each data sample.

Figure 4a visualizes the ultimate 10 features and their Pearson correlation matrix. We abbreviate the subscript "interstice" as "is"; and for several distance interstice features, the subscript "dist" in $d_{is\text{-}dist}$ indicates that the nearest-neighbors are determined by a cutoff distance rather than by the default Voronoi tessellation. The 10 features exhibit low Pearson correlation coefficient (the maximum is 0.63, Fig. 4a). Interestingly, we find that 9 out of the 10 survived features are describing interstice distribution in the medium-range (i.e. with "MRO" in the feature name). This again suggests that MRO contributes greatly to the decision making. According to the feature importance, $MRO_{mean}$ Std($V_{is}$) and $MRO_{mean}$ Std($d_{is\text{-}dist}$) are the most important features in the L-$E_{act}$ and H-$E_{act}$ tasks, respectively (Fig. 4b). These two metrics are evaluating the average variation of the tetrahedron volume interstice and bond distance interstice at the medium-range around an atom. This emphasizes the importance of local structure anisotropy, persisting to the medium-range, to the glass property. For the L-$E_{act}$ task, $MRO_{mean}$ Std($V_{is}$) stands out with a very high importance, and for the H-$E_{act}$ task, the feature importances distribute more evenly.

We then select typical hard and soft Cu (Zr) atoms and show the distribution of tetrahedron volume interstice, $V_{is}$, and bond distance interstice, $d_{is\text{-}dist}$, in their local environment to demonstrate the inherent structural contrast between the hard and soft atoms. Typical atoms with high $E_{act}$ (~2.9 eV) and low $E_{act}$ (~0.7 eV) are selected, and the red and purple histograms show the spread of interstices, $V_{is}$ and $d_{is\text{-}dist}$, present in the coordination polyhedron (SRO) and in the neighboring clusters (MRO), respectively (Fig. 5a and 5b). We find that the $V_{is}$ and $d_{is\text{-}dist}$ distributions in the SRO of the high $E_{act}$ atoms (Fig. 5a) are distinctly more centered than that in the low $E_{act}$ ones (Fig. 5b). For the low $E_{act}$ atoms, there often exist some tetrahedra or bond segments that have very low or high content of interstice. This would lower the stability of local environment and propel the atom to respond to thermal excitation. Remarkably, this trend persists to the medium-range (purple histograms). As quantified by Fig. 4b, the MRO interstice distribution is even more important than the SRO ones. The sharp contrast in the interstice distribution illustrates the foundation of our ML success in distinguishing the characteristic atoms.

Next, we use principal component analysis (PCA)[43] to project the information in the high-dimensional feature space ($R^{10}$, ten features in Fig. 4) into a low-dimensional space ($R^2$) to visualize the inherent data structure of the site environment features (Fig. 5c). PCA is a dimensionality reduction method that uses orthogonal transformation to reduce possibly correlated features to uncorrelated variables with key information preserved, and is totally unsupervised (with no use of class labels and does not need training)[43]. From Fig. 5c, we see that the high $E_{act}$ and low $E_{act}$ atoms do tend to reside in very different regions. Back to the above supervised ML results, strong structural contrast in both the hard and soft ends is also revealed (Figs. 2 and 3). Here, both supervised and unsupervised analyses suggest a highly inhomogeneous MG structure, with distinctive hard (or say solid-like) and soft (liquid-like) atoms dissolved inside.

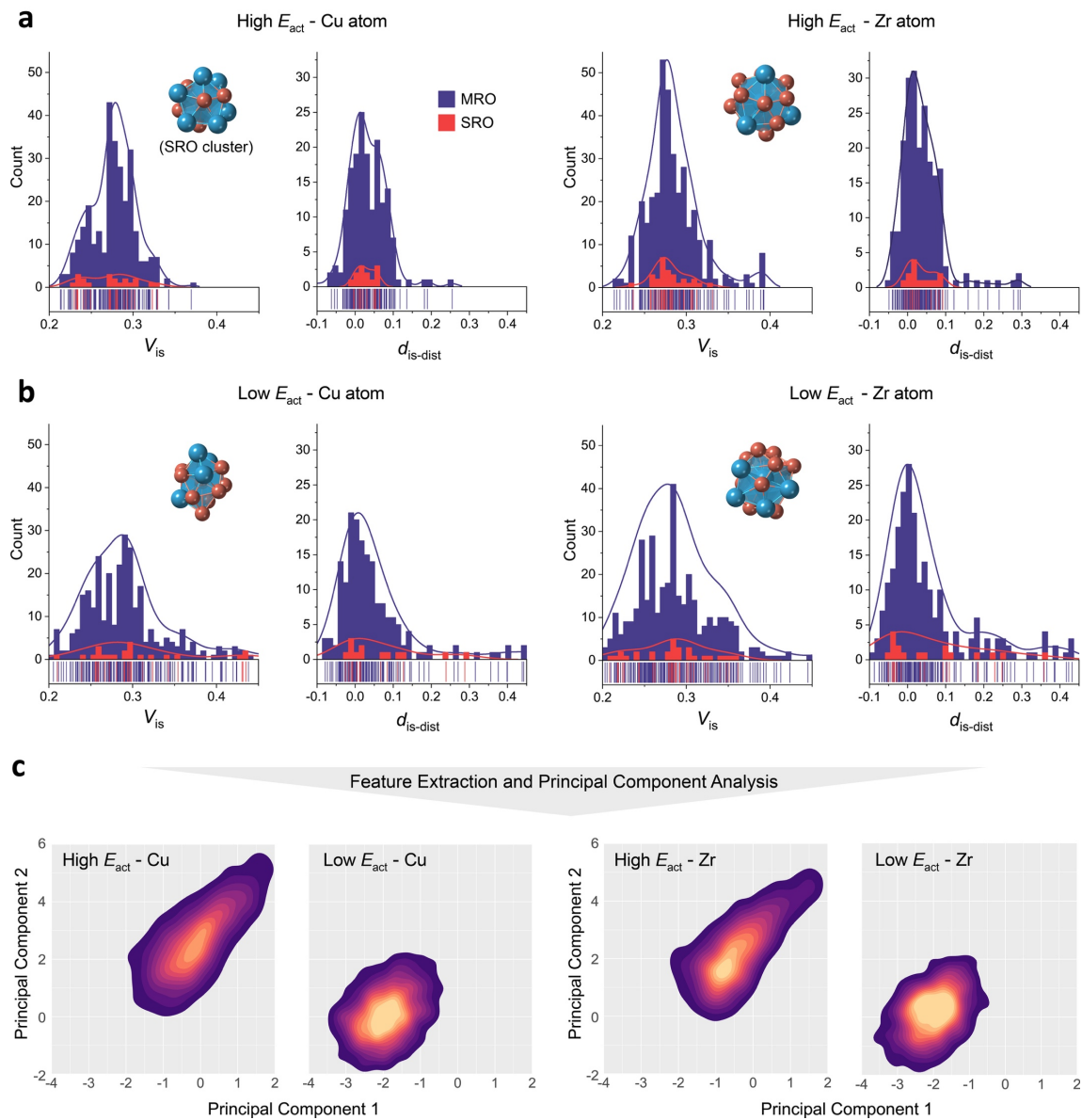

**Figure 5 | Structural contrast between the high and low $E_{act}$ atoms**. (a) Distribution of tetrahedron volume interstice $V_{is}$ and bond distance interstice $d_{is\text{-}dist}$ of representative Cu and Zr atoms with high $E_{act}$ = ~2.95 eV. The red and purple histogram shows the spread of interstices in the coordination cluster (SRO) and that in the neighboring clusters (MRO), respectively. The SRO coordination clusters are also visualized in the inset, with the larger blue spheres for Zr and smaller orange spheres for Cu. (b) Distribution of $V_{is}$ and $d_{is\text{-}dist}$ of representative Cu and Zr atoms with low $E_{act}$ = ~0.71 eV. (c) Unsupervised principal component analysis (PCA) to reduce the original high-dimensional interstice feature space to a two-dimensional space.

**Transferability to identifying shear transformation propensity**

As mentioned earlier, in addition to thermally activated events, another important type of elementary excitation is the local shear transformation activated by stress[44-49]. The low-stress-resistance units are usually referred to as shear transformation zones (STZs). As discussed in the Introduction, the thermal- and stress-activated excitations can both be interpreted in the framework of β processes, however, the atomic-specific response can vary, due to the different characteristics of stimulus source (uniform vs non-uniform, protocol-independent vs dependent). This prompts us to ask: how would our ML models trained for predicting the thermal excitation propensity perform, when they are used to identify STZs? Is it possible for the models to work well when transferring to such a different task?

This "between-task" test is challenging in several ways: i) STZs and $E_{act}$ are basically different properties, stimulated by different stimuli and thus yielding different data; ii) the features considered important for predicting $E_{act}$ may not be optimal for identifying STZs. The ii) point is very likely, as in a previous work using the interstice features to identify STZs in MGs, only ~50% of the most important features were MRO features[19], much lower than the ~90% in the $E_{act}$ case (Fig. 4). Driven by this question, we simulate athermal quasi-static (AQS) shear deformation of a typical $Cu_{64}Zr_{36}$ – $10^9$ K/s glass (Methods). We calculate the interstice features of each atom and apply the model trained from the L-$E_{act}$ problem (which focuses on the soft end) to derive the probability estimate $p_l$ of each atom. Intuitively, as $p_l$ is in positive correlation with the tendency of an atom to be easily activated by the thermal stimulus (Fig. 2 and Fig. 3), it may positively correlate with the susceptibility of atom to be activated by stress as well.

We calculate the non-affine displacement ($D^2_{min}$) relative to undeformed state, at 4.0% shear strain, as an indicator of the plastic susceptibility of each atom. The correlation between $D^2_{min}$ and $p_l$ is presented in Fig. 6a. Given the "long-tail" distribution of $D^2_{min}$, box plots are used to present the correlation. Box plots are useful in such skewed distributions, with the median (a line in the interior of box), 25% and 75% quantile (lower and upper ends of box), 1.5 times the inter-quartile range (whiskers extending outside box) as well as outliers (points outside the whiskers) cleared marked. The left figure in Fig. 6a shows the complete box plot, and some outliers extend so widely that the box section is squeezed. We then highlight the squeezed section, which constitutes the vast majority of data, in the right figure of Fig. 6a. A positive correlation between $p_l$ and $D^2_{min}$ is clearly observed, evidencing our assumption that these two types of activations could have some similar structural origins. As another quantitative test, we use $p_l$ to try classifying STZs with the largest 5% $D^2_{min}$ from the rest of the glass, similar to the setting of the L-$E_{act}$ task.

We vary the threshold of $p_l$ in designating the positive/negative classes in this STZ task, calculate the TPRs and FPRs and derive the ROC curve in Fig. 6b. The area under the ROC curve, AUC-ROC, is 0.810, which is a very reasonable score for such a transferring test. This quantitative test provides additional support to the feasibility of this "between-task" generalization.

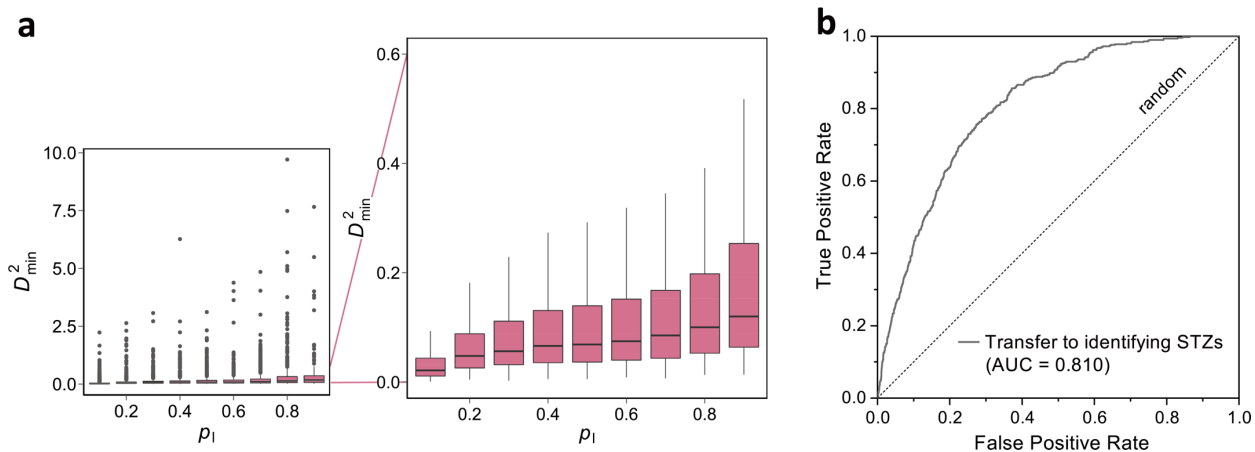

**Figure 6 | Transferring ML model to identifying STZs.** (a) Correlation between the probability estimates, $p_l$, from the L-$E_{act}$ model and the non-affine displacement ($D^2_{min}$) at strain 4.0% of a $Cu_{64}Zr_{36}$ – $10^9$ K/s glass. In the box plots, ends of box spans from 25 to 75% percentile, black line in box represents median, whiskers show 1.5 times the inter-quartile range, and points outside the whiskers show outliers. Outliers are marked in the left plot and are removed in the right plot for clarity. (b) Receiver operating characteristic (ROC) curve and area under curve (AUC) if using the $p_l$ to identify STZs with the largest 5% $D^2_{min}$.

As discussed in the Introduction, the accuracy of STZ recognition (for example, Ref. 18 and 19) is usually lower than that of identifying the thermally activated atoms (Ref. 17 and this work), especially when using the same feature representation (namely radial symmetry functions[17,18] or interstice distribution features[19]). There are several factors that can cause this performance difference. One is the increased internal data noise of the STZ data, if the data is collected from a single loading condition. As discussed in the Introduction, stress-activated plastic heterogeneity is quite sensitive to the loading conditions such as loading mode and direction[25,26]; thus, if using data from a single loading condition, non-trivial noise would be introduced in the collected data. For the thermal activation data, as used in this work, the absence of non-uniform stress eliminates the loading-related noises, and probing sufficient elementary ART events can guarantee a well-converged $E_{act}$ to indicate the susceptibility to thermal excitation. In addition, upon deformation, the activation of STZ proceeds in a progressive way, that is, not all soft atoms will move in a strain step; therefore, it usually requires a relatively large strain to collect sufficient plastic events. However, this can introduce more cascade activation events to reduce the controllability of the initial undeformed structure, and the existence of long-range elastic field in the process of deformation would also increase the length scale of plastic heterogeneity, making it beyond the scope of SRO and MRO that can be described by the structural representation.

## DISCUSSION

For the ML exploration of atomic-level structure-property relationships in amorphous alloys, a goal of common pursuit is developing novel structural representation and machine learning scheme. This paper, instead, focuses on another important aspect – finding of a suitable target property, with minimized data noises, to convincingly test the power of ML in correlating the structure with the property. Through what has been presented above, we have demonstrated that the thermally activated elementary excitation is an excellent choice in this regard. Compared with previous ML models on shear transformations in glasses, the merits of our present success on thermally activated events in MGs are multifold:

i) We reached an unprecedentedly high accuracy for ML prediction of elementary excitation in MGs. In this work, ML can accurately identify atoms with the highest 5% and lowest 5% thermal activation energy in a dataset merged from six different MGs, reaching an AUC-ROC of 0.942 and 0.888, respectively. These scores are significantly higher than that achieved in predicting the propensity of shear transformation. As discussed, this is mainly because the thermal activation does not suffer from the effect of non-uniform, oriented stress[25,26], and can reduce the data noises by well-converged exploration of elementary excitations. The importance of noise reduction also has implications for constructing high-fidelity glass datasets in the future.

ii) Our ML model is able to link with both local favored and unfavored structural motifs, rather than only identifying the latter as in previous ML literature[17-21]. This is aided by the explicit and sufficient ART perturbation tests around each atom, and the Gaussian-like distribution of thermal activation energetics that gives sufficient resolution to both the soft and hard ends.

iii) We have demonstrated that the data from multiple compositions or processing histories can be combined to connect with underlying structural signatures. This results from the comparable magnitude/range of activation barriers, for different compositions and processing histories in the same MG system. Such treatment can notably increase the variety of local environments surveyed, and allows for structure-property relation mining in more general terms.

iv) Our analysis provides a repertoire of descriptors that are essential to the ML decision. We demonstrate how the ML models make decisions based on the interstice features and interpret why these features work in representing the inherent structural contrast in MGs. Our data-centric results also highlight the importance of MRO features in determining the activation heterogeneity that has implications on the underlying glass physics.

v) We have conducted a quantitative "between-task" transferring test that successfully transfers the model fitted for pinpointing the low thermal activation energy atoms to identifying STZs upon AQS shear deformation. This success points to some common structural origins of the thermal-activated and stress-activated β processes.

Taken together, these advances underscore the structural impact on the β processes and their heterogeneity, and the insights shed light on the role of β processes as a basic unit event underlying a variety of properties of MGs[10-12], including local plastic deformation[13,14], atomic hoping mediating diffusion[15], and structural relaxation/rejuvenation[16]. Our discovery, enabled by the well-designed site environment representation and dedicated ML models, is very useful and important as a step forward in establishing a concrete structure-property relationship for MGs. We have made our MG configurations and thermal activation energy data public in figshare with the DOI of https://doi.org/10.6084/m9.figshare.12485795, which could serve as a valuable benchmark for future ML studies in MG research.

## METHODS

**MG samples preparation by MD simulation**

Molecular dynamics (MD) simulations using LAMMPS[50] have been employed to prepare and analyze the Cu-Zr metallic glass models, using a set of optimized embedded-atom-method (EAM) potentials[51]. $Cu_{64}Zr_{36}$, $Cu_{50}Zr_{50}$ and $Cu_{80}Zr_{20}$ samples containing 10,000 or 5,000 atoms (if 5,000, we will prepare two different samples at the same processing condition) were quenched to room temperature (300 K) from equilibrium liquids above the corresponding melting points. The quenching was performed at a rate of $10^9 - 10^{12}$ K/s, as marked in Fig. 1b, using a Nose-Hoover thermostat with zero external pressure. Periodic boundary conditions (PBC) were applied in all three directions during MD simulation[52]. The timestep is 1 fs.

**Activation-relaxation technique (ART)**

Initial perturbations in ART were introduced by applying random displacement on a small group of atoms (an atom and its nearest-neighbors)[27,28]. The magnitude of the displacement was fixed, while the direction was randomly chosen. When the curvature of the PEL was found to overcome the chosen threshold, the system was pushed towards the saddle point using the Lanczos algorithm. The saddle point is considered to be found when the overall force of the total system is below 0.01 eV/Å. The corresponding activation energy is thus the difference between the saddle point energy and the initial state energy. The search is performed using ART nouveau package[27,28,53]. For each group of atoms, we employed ~50 successful ART searches with different random perturbation directions.

**Flexibility volume and atomic shear moduli**

The flexibility volume $V_{\text{flex},i}$ of atom $i$ is defined as[41]:

$$V_{\text{flex},i} = \left\langle \left(x_i(t) - \bar{x}_i\right)^2 \right\rangle \times V_i^{1/3} \tag{3}$$

where $\bar{x}_i$ and $x_i(t)$ are the equilibrium position and instantaneous position at time $t$ of the atom $i$, and $V_i$ is the corresponding atomic volume. The calculation was obtained on short time scales when the mean square displacement is flat with time and contains the vibrational but not the diffusional contribution. Each sample was kept at equilibrium under a microcanonical ensemble (NVE) at room temperature for the calculation, which was taken over 100 independent runs, all starting from the same configuration but with momenta assigned randomly from the appropriate Maxwell-Boltzmann distribution.

Atomic shear moduli at room temperature were evaluated using the fluctuation method. For a canonical (NVT) ensemble, elastic constants can be calculated as the sum of three contributions:

$$C_{ijkl}^T = C_{ijkl}^I + C_{ijkl}^{II} + C_{ijkl}^{III} \tag{4}$$

where the superscript *I, II* and *III* represents the fluctuation, kinetic contribution and the Born term, respectively (see Ref. 42 for more details). To reduce the statistical error in our simulated samples, the average atomic shear modulus ($G$) was evaluated as

$$G = \frac{C_{44} + C_{55} + C_{66}}{3} \tag{5}$$

The local elastic moduli tensor is computed at the coarse-grained scale using the average atomic shear moduli of center atom and its nearest neighbors.

**Athermal quasi-static (AQS) simulation**

We employ the athermal quasi-static (AQS) mode to simulate the shear deformation of the glass[54]. On each deformation step, an affine strain of $10^{-4}$ is imposed along the +xy direction, followed by an energy minimization using the conjugate-gradient method. Initial configuration is the inherent structure of the equilibrated glass sample. The simulations were conducted using LAMMPS[50] and periodic boundary conditions (PBC) were applied in all three directions. The plastic events were monitored using the non-affine displacement ($D^2_{min}$)[45]. This is done by tracking the atomic strain of each atom during deformation, and dissociating the strain into the best affine fit and the non-affine residue.

# DATA AND CODE AVAILABILITY

The datasets used in this work have been made public in figshare with the DOI of https://doi.org/10.6084/m9.figshare.12485795. The codes for the interstice representation can be publicly accessed in *amlearn*[19] (https://github.com/Qi-max/amlearn) and *matminer*[30] (https://github.com/hackingmaterials/matminer).

# REFERENCES

1. Greer, A. L. Metallic Glasses. in *Physical Metallurgy: Fifth Edition* 305–385 (Elsevier, 2014).
2. Schroers, J. Bulk metallic glasses. *Phys. Today.* **66**, 32-37 (2013).
3. Egami, T. Atomic level stresses. *Prog. Mater. Sci.* **56,** 637-653 (2011).
4. Hirata, A. *et al.* Direct observation of local atomic order in a metallic glass. *Nat. Mater.* **10**, 28–33 (2011).
5. Cheng, Y. Q. & Ma, E. Atomic-level structure and structure-property relationship in metallic glasses. *Prog. Mater. Sci.* **56**, 379–473 (2011).
6. Goldstein, M. Viscous Liquids and the Glass Transition: A Potential Energy Barrier Picture. *J. Chem. Phys.* **51**, 3728–3739 (1969).
7. Debenedetti, P. G. & Stillinger, F. H. Supercooled liquids and the glass transition. *Nature* **410**, 259–267 (2001).

8. Wales, D. J. A Microscopic Basis for the Global Appearance of Energy Landscapes. *Science* **293**, 2067–2070 (2001).

9. Johari, G. P. & Goldstein, M. Viscous Liquids and the Glass Transition. II. Secondary Relaxations in Glasses of Rigid Molecules. *J. Chem. Phys.* **53**, 2372–2388 (1970).

10. Yu, H.-B., Wang, W.-H. & Samwer, K. The β relaxation in metallic glasses: an overview. *Mater. Today* **16**, 183–191 (2013).

11. Qiao, J. C. & Pelletier, J. M. Dynamic Mechanical Relaxation in Bulk Metallic Glasses: A Review. *J. Mat. Sci. Technol.* **30**, 523-545 (2014).

12. Yu, H.-B., Richert, R. & Samwer, K. Structural rearrangements governing Johari-Goldstein relaxations in metallic glasses. *Sci. Adv.* **3**, e1701577 (2017).

13. Fan, Y., Iwashita, T. & Egami, T. How thermally activated deformation starts in metallic glass, *Nat. Commun.,* **5**, 5083 (2014).

14. Wang, Z., Sun, B. A., Bai, H. Y. & Wang, W. H. Evolution of hidden localized flow during glass-to-liquid transition in metallic glass. *Nat. Commun.* **5**, 5823 (2014).

15. Yu, H. B., Samwer, K., Wu, Y. & Wang, W. H. Correlation between β relaxation and self-diffusion of the smallest consituting atoms in metalllic glasses. *Phys. Rev. Lett.* **109**, 095508 (2012).

16. Zhu, F. *et al.* Intrinsic correlation between β-relaxation and spatial heterogeneity in a metallic glass. *Nat. Commun.* **7**, 11516 (2016).

17. Schoenholz, S. S., Cubuk, E. D., Sussman, D. M., Kaxiras, E. & Liu, A. J. A structural approach to relaxation in glassy liquids. *Nat. Phys.* **12**, 469–471 (2016).

18. Cubuk, E. D. *et al.* Identifying structural flow defects in disordered solids using machine-learning methods. *Phys. Rev. Lett.* **114**, 108001 (2015).

19. Wang, Q. & Jain, A. A transferable machine-learning framework linking interstice distribution and plastic heterogeneity in metallic glasses. *Nat. Commun.* **10**, 5537 (2019).

20. Cubuk, E. D. *et al*. Structure-property relationships from universal signatures of plasticity in disordered solids. *Science*. **358**, 1033-1037 (2017).

21. Harrington, M., Liu, A. J. & Durian, D. J. Machine learning characterization of structural defects in amorphous packings of dimers and ellipses. *Phys. Rev. E.* **99**, 022903 (2019).

22. Sussman, D. M., Schoenholz, S. S., Cubuk, E. D. & Liu, A. J. Disconnecting structure and dynamics in glassy thin films. *Proc. Natl Acad. Sci.* **114**, 10601–10605 (2017).

23. Ma, Xiaoguang *et al*. Heterogeneous Activation, Local Structure, and Softness in Supercooled Colloidal Liquids. *Phys. Rev. Lett.* **122**, 28001 (2019).

24. Landes, F. P *et al*. Attractive versus truncated repulsive supercooled liquids: The dynamics is encoded in the pair correlation function. *Phys. Rev. E* **101**, 010602 (2020).

25. Barbot, A. *et al*. Local yield stress statistics in model amorphous solids. *Phys. Rev. E.* **97**, 33001 (2018).

26. Schwartzman-Nowik, Z., Lerner, E. & Bouchbinder, E. Anisotropic structural predictor in glassy materials. *Phys. Rev. E.* **99**, 60601 (2019).

27. Barkema, G. T. & Mousseau, N. Event-Based Relaxation of Continuous Disordered Systems. *Phys. Rev. Lett.* **77**, 4358–4361 (1996).

28. Rodney, D. & Schuh, C. Distribution of Thermally Activated Plastic Events in a Flowing Glass. *Phys. Rev. Lett.* **102**, 235503 (2009).

29. Lee, M., Lee, C. M., Lee, K. R., Ma, E. & Lee, J. C. Networked interpenetrating connections of icosahedra: Effects on shear transformations in metallic glass. *Acta Mater.* **59**, 159–170 (2011).

30. Ward, L. *et al.* Matminer: An open source toolkit for materials data mining. *Comput. Mater. Sci.* **152**, 60–69 (2018).

31. Chen, T. & Guestrin, C. XGBoost: A Scalable Tree Boosting System. in *Proceedings of the 22nd ACM SIGKDD International Conference on Knowledge Discovery and Data Mining* 785–794 (Association for Computing Machinery, 2016). doi:10.1145/2939672.2939785

32. Bradley, A. P. The use of the area under the ROC curve in the evaluation of machine learning algorithms. *Pattern Recognit.* **30**, 1145–1159 (1997).

33. Zadrozny, B. & Elkan, C. Obtaining calibrated probability estimates from decision trees and naive Bayesian classifiers. in *Proceedings of the 18th International Conference on Machine Learning* (2001).

34. Cortes, C. & Vapnik, V. Support-vector networks. *Mach. Learn.* **20**, 273–297 (1995).

35. Okabe, A., Boots, B., Sugihara, K. & Chiu, S. N. *Spatial Tesselations*. *Concepts and Applications of Voronoi Diagrams* (2009). doi:10.1002/0471721182.scard

36. Ding, J., Patinet, S., Falk, M. L., Cheng, Y. & Ma, E. Soft spots and their structural signature in a metallic glass. *Proc. Natl. Acad. Sci.* **111**, 14052--14056 (2014).

37. Ding, J., Cheng, Y. Q. & Ma, E. Full icosahedra dominate local order in Cu64Zr34 metallic glass and supercooled liquid. *Acta Mater.* **69**, 343–354 (2014)25.

38. Steinhardt, P. J., Nelson, D. R. & Ronchetti, M. Bond-orientational order in liquids and glasses. *Phys. Rev. B* **28**, 784–805 (1983).

39. Behler, J. & Parrinello, M. Generalized Neural-Network Representation of High-Dimensional Potential-Energy Surfaces. *Phys. Rev. Lett.* **98**, 146401 (2007).

40. Bartók, A. P., Kondor, R. & Csányi, G. On representing chemical environments. *Phys. Rev. B* **87**, 184115 (2013).

41. Ding, J. *et al.* Universal structural parameter to quantitatively predict metallic glass properties. *Nat. Commun.* **7**, 13733 (2016).

42. Cheng, Y. Q. & Ma, E. Configurational dependence of elastic modulus of metallic glass. *Phys. Rev. B* **80**, 64104 (2009).

43. Tipping, M. E. & Bishop, C. M. Probabilistic Principal Component Analysis. *J. R. Stat. Soc. Ser. B (Statistical Methodol.* **61**, 611–622 (1999).

44. Argon AS. Plastic deformation in metallic glasses, *Acta Metall.* **27**, 47 (1979)

45. Falk, M. L. & Langer, J. S. Dynamics of viscoplastic deformation in amorphous solids. *Phys. Rev. E.* **57**, 7192–7205 (1998).

46. Tsamados, M., Tanguy, A., Goldenberg, C. & Barrat, J.L. Local elasticity map and plasticity in a model Lennard-Jones glass, *Phys. Rev. E* **80**, 026112 (2009).

47. Greer, A. L., Cheng, Y. Q. & Ma, E. Shear bands in metallic glasses. *Mater. Sci. Eng. R Reports* **74**, 71–132 (2013).

48. Hufnagel, T. C., Schuh, C. A. & Falk, M. L. Deformation of metallic glasses: Recent developments in theory, simulations, and experiments. *Acta Mater.* **109**, 375–393 (2016).

49. Wisitsorasak, A. & Wolynes, P. G. Dynamical theory of shear bands in structural glasses. *Proc. Natl. Acad. Sci.* **114**, 1287–1292 (2017).

50. Plimpton, S. Fast parallel algorithms for short-range molecular dynamics. *J. Comput. Phys.* **117**, 1–19 (1995).

51. Cheng, Y. Q., Ma, E. & Sheng, H. W. Atomic Level Structure in Multicomponent Bulk Metallic Glass. *Phys. Rev. Lett.* **102**, 245501 (2009).

52. Allen, M.P. & Tildesley, D.J. Computer Simulation of liquids (Clarendon Press, Oxford) (1987)

53. Marinica, M.C., Willaime, F. & Mousseau, N. Energy landscape of small clusters of self-interstitial dumbbells in iron. *Phys. Rev. B*. **83**, 9, (2011).

54. Maloney, C. E., & Lemaître, A. Amorphous systems in athermal, quasistatic shear. *Phys. Rev. E.* **74**, 016118 (2006).

## ACKNOWLEDGEMENTS

Q.W. and E.M. are supported at JHU by U.S. Department of Energy (DOE), DOE-BES-DMSE, under grant DE-FG02-19ER46056. J.D. acknowledges the Young Talent Startup Program of Xi'an Jiaotong University.

## AUTHOR CONTRIBUTIONS

Q.W. and J.D. initiated the plan for this study. Q.W. designed and analyzed the machine learning models. J.D. conducted the ART simulations and the flexibility volume and shear moduli analyses. Q.W., J.D. and E.M. discussed the results and wrote the manuscript.

## COMPETING INTERESTS

The authors declare no competing interests.

# Predicting the propensity for thermally activated β events in metallic glasses via interpretable machine learning

Qi Wang[1*], Jun Ding[2*], Evan Ma[1]

[1] Department of Materials Science and Engineering, Johns Hopkins University, Baltimore, Maryland 21218, United States

[2] Center for Advancing Materials Performance from the Nanoscale (CAMP-Nano), State Key Laboratory for Mechanical Behavior of Materials, Xi'an Jiaotong University, Xi'an 710049, China

* Correspondence to qiwang.mse@gmail.com (Q. W.), dingsn@xjtu.edu.cn (J. D.)

## 1. ML performance employing other feature representations

**1) <0,0,12,0,0> + <0,0,12,4,0> (baseline 1)**

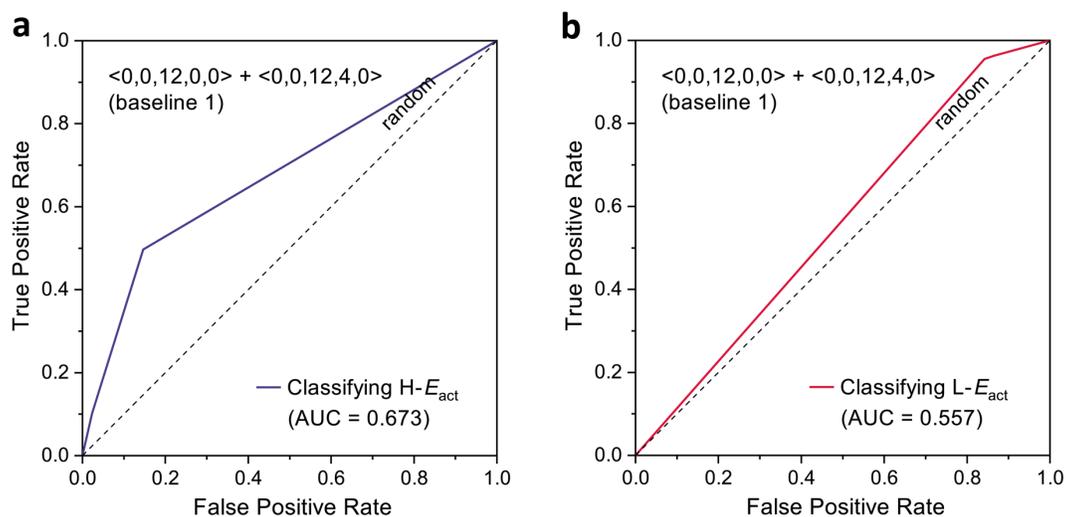

**Supplementary Fig. 1** | Receiver operating characteristic (ROC) curve of baseline model 1 fitted from features indicating whether the nearest-neighbors of an atom form a <0,0,12,0,0> or <0,0,12,4,0> Voronoi polyhedron or not. (a) Classifying the highest 5% $E_{act}$ atoms of the combined dataset merged from six MGs (Fig. 1b); (b) Classifying the lowest 5% $E_{act}$ atoms of the combined dataset.

## 2) Voronoi index (baseline 2)

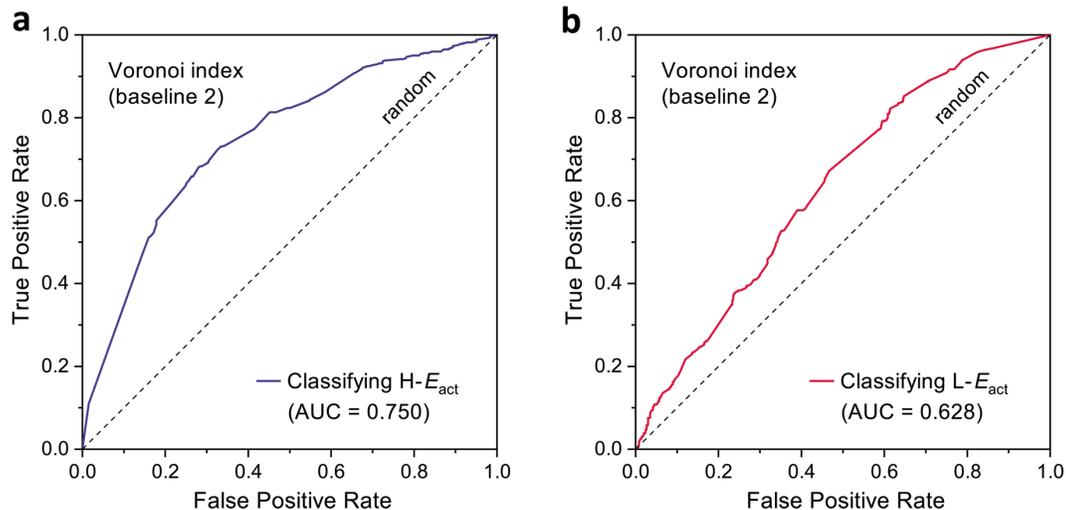

**Supplementary Fig. 2** | Receiver operating characteristic (ROC) curve of baseline model 2 fitted from the five integer Voronoi indices ($n_3$, $n_4$, $n_5$, $n_6$, and $n_{>6}$), where $n_x$ represents the number of $x$-edged facets in the Voronoi polyhedron. (a) Classifying the highest 5% $E_{act}$ atoms of the combined dataset merged from six MGs (Fig. 1b); (b) Classifying the lowest 5% $E_{act}$ atoms of the combined dataset.

## 3) A group of SRO features

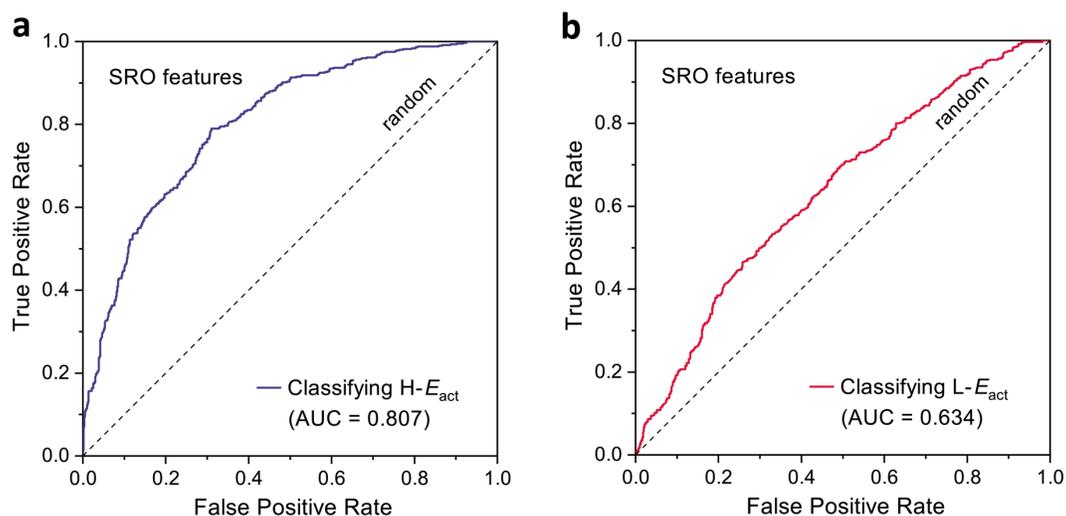

**Supplementary Fig. 3** | Receiver operating characteristic (ROC) curve of the ML model fitted from a group of SRO features (see Table 1 and related content for details). (a) Classifying the highest 5% $E_{act}$ atoms of the combined dataset merged from six MGs (Fig. 1b); (b) Classifying the lowest 5% $E_{act}$ atoms of the combined dataset.

## 4) SRO + MRO features

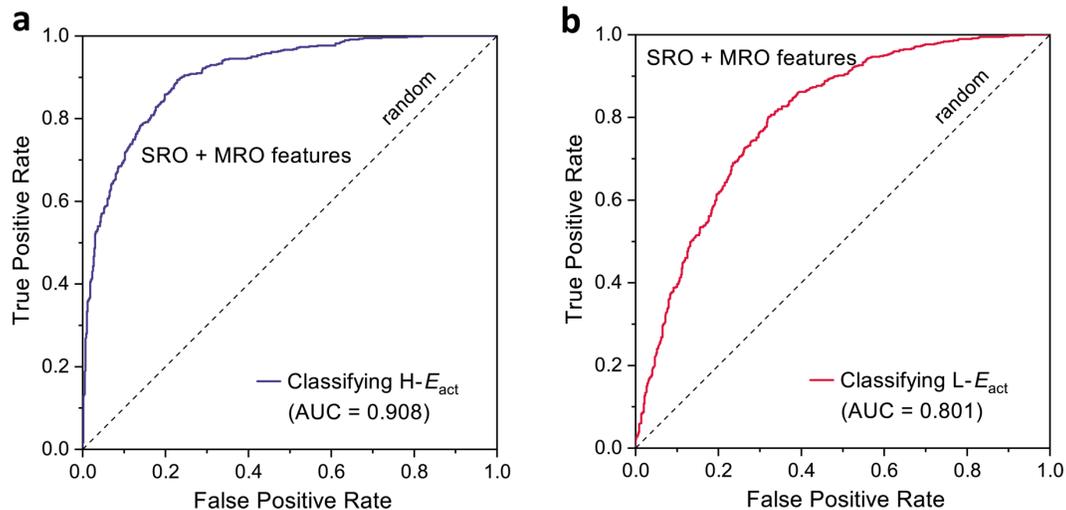

**Supplementary Fig. 4** | Receiver operating characteristic (ROC) curve of the ML model fitted from SRO features and coarse-grained MRO features derived by tasking statistics of SRO features in the neighboring shell. (a) Classifying the highest 5% $E_{act}$ atoms of the combined dataset merged from six MGs (Fig. 1b); (b) Classifying the lowest 5% $E_{act}$ atoms of the combined dataset.

## 5) Radial symmetry functions

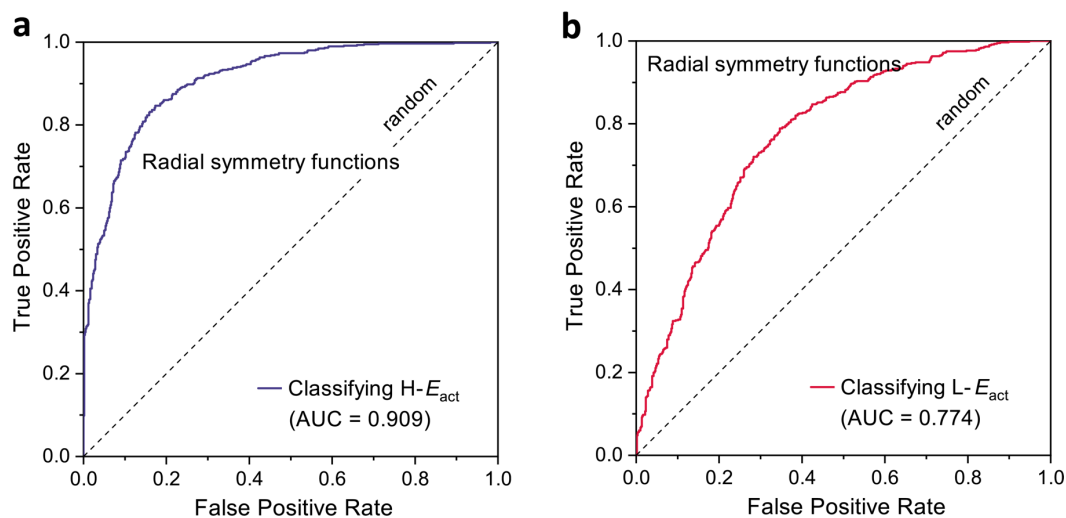

**Supplementary Fig. 5** | Receiver operating characteristic (ROC) curve of the ML model fitted from a set of radial symmetry functions. (a) Classifying the highest 5% $E_{act}$ atoms of the combined dataset merged from six MGs (Fig. 1b); (b) Classifying the lowest 5% $E_{act}$ atoms of the combined dataset.

## 6) Flexibility volume, $V_{flex}$

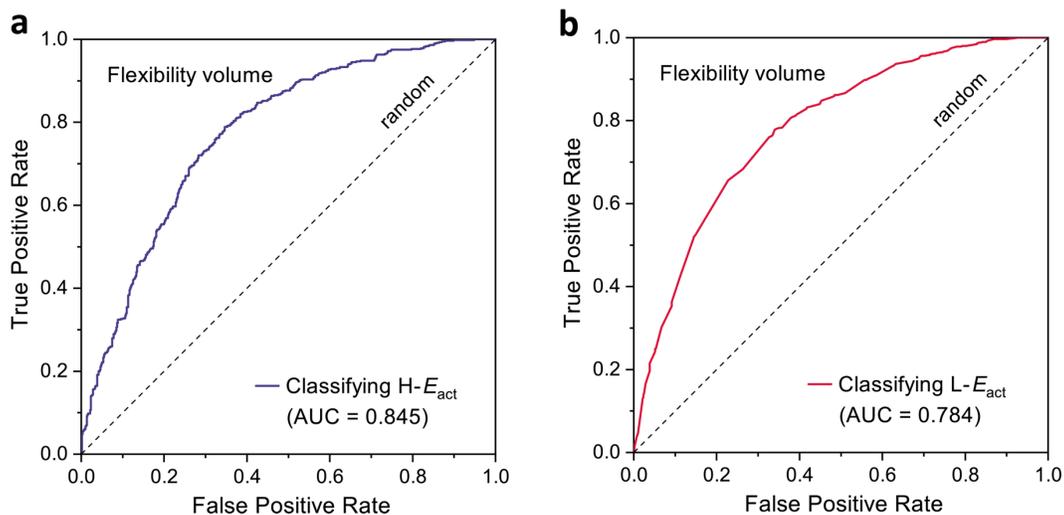

**Supplementary Fig. 6** | Receiver operating characteristic (ROC) curve of the ML model fitted from the flexibility volume, $V_{flex}$. (a) Classifying the highest 5% $E_{act}$ atoms of the combined dataset merged from six MGs (Fig. 1b); (b) Classifying the lowest 5% $E_{act}$ atoms of the combined dataset.

## 7) Atomic shear moduli, $G$

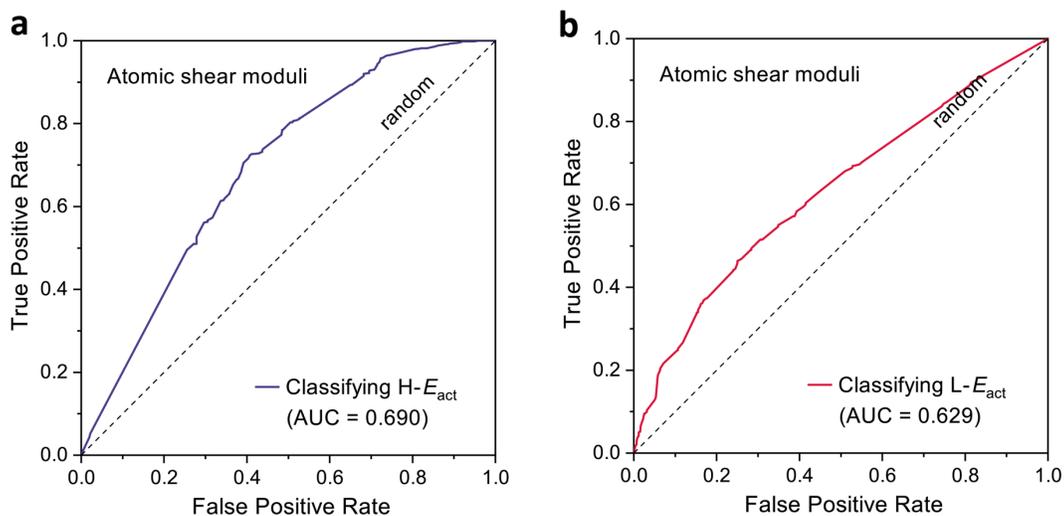

**Supplementary Fig. 7** | Receiver operating characteristic (ROC) curve of the ML model fitted from the atomic shear moduli, $G$. (a) Classifying the highest 5% $E_{act}$ atoms of the combined dataset merged from six MGs (Fig. 1b); (b) Classifying the lowest 5% $E_{act}$ atoms of the combined dataset.

## 8) Coarse-grained *G*

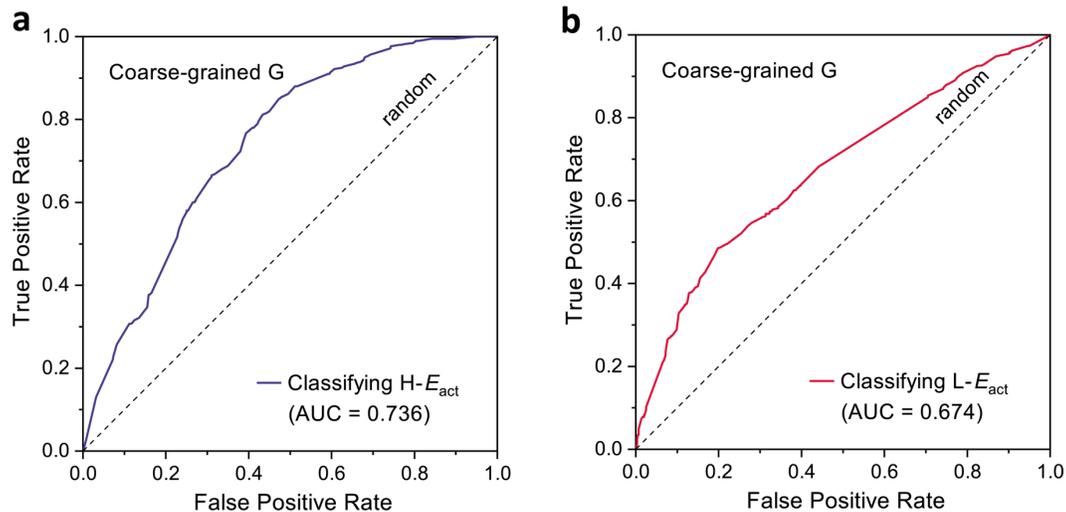

**Supplementary Fig. 8** | Receiver operating characteristic (ROC) curve of the ML model fitted from the coarse-grained atomic shear moduli, *G*. (a) Classifying the highest 5% $E_{act}$ atoms of the combined dataset merged from six MGs (Fig. 1b); (b) Classifying the lowest 5% $E_{act}$ atoms of the combined dataset.